\documentstyle[latexsym,amssymb,aps,floats,eqsecnum,
preprint,amsfonts]{revtex}
\def\laq{\raise 0.4ex\hbox{$<$}\kern -0.8em\lower 0.62 ex\hbox{$\sim$}}
\def\gaq{\raise 0.4ex\hbox{$>$}\kern -0.7em\lower 0.62 ex\hbox{$\sim$}}
\input epsf
\begin{document}

\bibliographystyle{unsrt}

\title{Gauge-invariant fluctuations of scalar branes}

\author{Massimo Giovannini
\footnote{Electronic address: 
Massimo.Giovannini@ipt.unil.ch}}

\address{{\it Institute of Theoretical Physics, 
University of Lausanne}}
\address{{\it BSP-1015 Dorigny, Lausanne, Switzerland}}

\maketitle

\begin{abstract} 
A generalization of the Bardeen formalism to the case 
of warped geometries is presented. The system determining the gauge-invariant
fluctuations of the metric induced by the scalar fluctuations 
of the brane is reduced to a set of Schr\"odinger-like equations for 
the Bardeen potentials and for the canonical normal 
modes of the scalar-tensor action. Scalar, vector and tensor modes 
of the geometry are classified according to four-dimensional 
Lorentz transformations.
While the tensor modes of the geometry live on the brane determining 
the corrections to Newton law, 
the scalar and and vector fluctuations  exhibit 
non normalizable zero modes and are, consequently, not localized on the brane.
The spectrum of the massive modes of the fluctuations 
is analyzed using supersymmetric quantum mechanics.
\end{abstract}
\vskip0.5pc
\centerline{Preprint Number: UNIL-IPT-01-07, May 2001 }
\vskip0.5pc
\noindent

\newpage
\renewcommand{\theequation}{1.\arabic{equation}}
\setcounter{equation}{0}
\section{Introduction}
Higher dimensional topological defects 
allow to investigate the properties and implications of warped 
geometries \cite{m2,ak,vis}. Fields of various spin can be 
localized on warped backgrounds. For instance, chiral fermions 
can be localized on a five-dimensional domain-wall solution \cite{m1}. 
Moreover, also gravitational interactions 
can be localized on five-dimensional walls \cite{rs,rs2}. For a recent review
 see \cite{rub}.

Suppose that a smooth domain-wall solution is used in order to 
localize fields. Such a geometry may be  generated, for 
instance, by a scalar potential 
$V(\varphi)$ with spontaneous breaking of $\varphi \rightarrow -\varphi$
symmetry \cite{m1}. Consider first the case where 
gravity is absent. In this case the scalar field has 
a normalizable zero mode \cite{m1}.

If gravity is consistently included
 the fluctuations of the scalar wall will induce 
fluctuations in the metric. It is then natural to ask if 
all the metric fluctuations are localized on the wall. There are 
two problems in this analysis. The first one is that
 metric fluctuations are coupled to the fluctuations of the 
wall. The second one is that  the coupled system 
of equations depends on the coordinate system.   
In order to discuss the metric fluctuations of scalar-tensor 
actions, the {\em Bardeen formalism} \cite{bar} has been very useful in 
four-dimensional backgrounds. The main idea is to parameterize 
the metric fluctuations by defining suitable gauge-invariant 
variables which do not change for infinitesimal coordinate 
transformations. The Bardeen formalism is also rather effective in order 
to identify the coordinate systems where the gauge-invariant 
variables take a simple form.  Having generalized the Bardeen 
formalism to the case of static five-dimensional warped geometries, 
the localization properties of metric 
fluctuations can be investigated. The standard lore is that 
if the zero mode of a given fluctuation is 
not normalizable, then it will not be localized on the brane and 
it will not affect the four-dimensional physics.
 
Explicit physical (thick) brane solutions have  been derived 
in different numbers of transverse dimensions. In five dimensions 
physical branes can be obtained using a scalar domain-wall where 
the scalar field depends upon the bulk radius \cite{kt1,kt2,gremm1,gremm2}. 
In \cite{free,free2} the stability of scalar domain walls (inspired by gauged 
supergravity theories) has been analyzed. 
In more than five dimensions, physical brane solutions including  also 
 background gauge fields have been recently analyzed \cite{gms}. 
Singular (scalar) brane solutions have been also 
analyzed from different points of view \cite{ctc}.
The present analysis deals with the case 
of  one transverse (bulk) coordinate. The proposed formalism can 
be however generalized to the case of higher dimensional 
transverse space.  

The generalization of the Bardeen 
formalism to the case of compact extra-dimensions has been addressed in a 
cosmological context.
The problem was to discuss the 
evolution of  metric fluctuations in cosmological models with 
compact extra-dimensions with \cite{mg1,mg2} and without \cite{ab} bulk 
scalars. In the present analysis the symmetry of the four-dimensional 
world is Lorentz invariance which will be used in order 
to classify the fluctuations of the metric \cite{t}. The approach 
discussed in the present paper is then different from the ones 
previously discussed, for cosmological applications, 
in \cite{mg1,mg2,ab} and in \cite{br}. 

The plan of this paper is the following. In Section II scalar-tensor 
models of physical branes will be briefly discussed. Emphasis will be given 
to the aspects relevant in the analysis of the perturbations.
In Section III the Bardeen formalism will be generalized to the case 
of warped metrics. Variables invariant under infinitesimal coordinate 
transformations will be discussed. Physical gauge choices will be proposed.
In Section IV the evolution for the scalar, tensor and vector modes 
of the geometry will be derived.  The system will be reduced to a 
set of (decoupled) second order differential equations. In Section 
V the zero-modes of the gauge-invariant variables will be discussed.
General conditions on the normalizability of the gravi-photon and 
gravi-scalar will be derived. Section VI contains some concluding remarks.
In the Appendix useful technical results have been collected.

\renewcommand{\theequation}{2.\arabic{equation}}
\setcounter{equation}{0}
\section{Background models} 

In five dimensions only one bulk coordinate 
is present and smooth domain wall configurations can be found
using the following scalar-tensor action:
\begin{equation}
S= \int d^{5}x \sqrt{|G|}\biggl[- \frac{R}{2\kappa} + \frac{1}{2} G^{A B} 
\partial_{A} \varphi \partial_{B} \varphi - V(\varphi)\biggr],
\label{totaction}
\end{equation}
whose related equations of motion are
\begin{eqnarray}
&& R_{A}^{B} - \frac{1}{2} \delta_{A}^{B} R = \kappa
 \biggl[\partial_{A}\varphi 
\partial^{B} \varphi - \delta_{A}^{B} \biggl( \frac{1}{2} G^{M N} \partial_{M} 
\varphi\partial_{N} \varphi - V(\varphi)\biggr)\biggr],
\label{e}\\
&& G^{A B} \nabla_{A} \nabla_{B} \varphi + 
\frac{\partial V}{\partial \varphi} =0.
\label{ph}
\end{eqnarray}
Notice that $\kappa = 8\pi G_{5} = 8\pi/M_{5}^3$. 
In order to facilitate the treatment of the fluctuations it is 
useful to write the five-dimensional metric in conformal coordinates, namely
\footnote{The conventions will be the following. Greek indices run over 
the four-dimensional space. Latin 
(uppercase) indices run over all the five dimensions.}:
\begin{equation}
ds^2 = {\overline G}_{A B} dx^A d x^B  = a^2(w) [dt^2 - d\vec{x}^2 - dw^2].
\label{m1}
\end{equation}
Choosing  now natural gravitational units $ 2 \kappa = 1$,
the explicit form of Eqs. (\ref{e})--(\ref{ph}) is:
\begin{eqnarray}
&& \varphi'' + 3 {\cal H} \varphi' - \frac{\partial V}{\partial\varphi} a^2 =0,
\label{s}\\
&& {\cal H}' + {\cal H}^2 = - \frac{1}{6} 
\biggl[ \frac{\varphi'^2 }{2} + V a^2 \biggr], 
\label{e1}\\
&&{\cal H}^2 = \frac{1}{12} \biggl[ \frac{\varphi'^2}{2} - V a^2\biggr].
\label{e2}
\end{eqnarray}
where
\begin{equation}
{\cal H} = \frac{a'}{a},
\end{equation}
and the prime denotes the derivation with respect to the bulk coordinate $w$.
Eqs. (\ref{s})--(\ref{e2}) are not independent. By combining two of them 
the third one can be obtained. This
 property can be used in order to solve the full system.

The equation for $\varphi$ can be obtained by combining Eqs. (\ref{e1}) 
and (\ref{e2}):
\begin{eqnarray}
&&\varphi'^2 = 6 ( {\cal H}^2 - {\cal H}'), 
\label{c1}\\
&& Va^2 = - 3 ( 3 {\cal H}^2 + {\cal H}').
\label{c2} 
\end{eqnarray}
Eqs. (\ref{c1}) and (\ref{c2}) will allow to simplify some of the equations
for the fluctuations.

Given a specific form of $a(w)$, Eq. (\ref{c1}) determines 
$\varphi(w)$ whereas, from  Eq. (\ref{c2}), $V(w)$ is obtained. 
Inserting $w(\varphi)$
into $V(w)$, the specific form of 
$V(\varphi)$ is found, once $a(w)$ is specified. 

Using this method of integration various exact solutions 
can be discussed \cite{kt1,kt2,gremm1,gremm2}. 
Most of the results reported in this paper 
will be general but in order to fix the ideas it is useful to 
have in mind the following regular geometry :
\begin{equation}
a(w)  = \frac{1}{\sqrt{ b^2 w^2 + 1}}, 
\label{s1}
\end{equation}
where $b$ is a real number.
Using Eq. (\ref{s1}) in Eq. (\ref{c1}) and integrating once the scalar field
corresponding to the geometry (\ref{s1}) is  
\begin{equation}
\varphi(w) = \sqrt{6} \arctan{b w}.
\label{s1a}
\end{equation}
Inserting $w(\varphi)$ into 
 Eq. (\ref{c2}) 
\begin{equation}
V(\varphi) = 3 b^2 \biggl[ 1 - 5 \sin^2 \frac{\varphi}{\sqrt{6}}\biggr].
\label{s1b}
\end{equation}
Eqs. (\ref{s1})--(\ref{s1b}) describe a thick domain wall solution. The 
metric associated with this field configuration is non-singular (i.e. 
no poles are present in the curvature invariants). 
In the non-conformal coordinate system the metric 
is determined recalling that $a(w) dw = dy$. Integrating 
once Eq. (\ref{s1}) 
\begin{equation}
a(y) = \frac{1}{\cosh{b y}}.
\end{equation}
Most of the results reported in the following sections will also
hold for singular domain-wall solutions \cite{ctc} 
where the scalar field is coupled to the action of a thin wall 
present for some finite value of the bulk coordinate.

\renewcommand{\theequation}{3.\arabic{equation}}
\setcounter{equation}{0}
\section{Fluctuations of the metric and of the bulk field} 
\subsection{Preliminaries}

The fluctuations of the metric (\ref{m1}) are coupled to 
the fluctuations of the bulk scalar $\varphi$, as it can be 
appreciated by writing down the equations for the 
fluctuations in their general form. Contracting once Eq. (\ref{e})
the background equations can be written as
\begin{eqnarray}
&& R_{A B} = \frac{1}{2} \partial_{A} \varphi \partial_{B} \varphi
- \frac{V}{3} G_{A B} ,
\nonumber\\
&& G^{A B} \biggl( \partial_{A} \partial_{B}\varphi
 - \Gamma^{C}_{A B} \partial_{C}
 \varphi \biggr) + \frac{\partial V}{\partial\varphi} \chi =0,
\end{eqnarray}
leading to the evolution equations of the fluctuations
\begin{eqnarray}
&&\delta R_{A B} = \frac{1}{2} \partial_{A} \varphi \partial_{B} \chi + 
\frac{1}{2} \partial_{A} \chi \partial_{B} \varphi - \frac{1}{3} 
\frac{\partial V}{\partial\varphi} \chi \overline{G}_{AB} - \frac{V}{3} 
\delta G_{A B},
\label{P1}\\
&& \delta G^{A B} \biggl( \partial_{A} \partial_{B} \varphi -
\overline{\Gamma}_{A B}^{C} \partial_{C} \varphi \biggr) 
+ \overline{G}^{A B} \biggl( \partial_{A} \partial_{B} \chi - 
\overline{\Gamma}_{A B}^{C} \partial_{C} \chi  - \delta 
\Gamma^{C}_{A B} \partial\varphi\biggr) 
+ \frac{\partial^2 V}{\partial \varphi^2}\chi=0.
\label{P2}
\end{eqnarray}
The metric and the scalar field have been separated  
into their background and perturbation parts:
\begin{eqnarray}
&& G_{AB}(x^{\mu}, w) = \overline{G}_{AB}(w) + \delta G_{AB}(x^{\mu},w),
\nonumber\\
&& \varphi(x^{\mu}, w) = \varphi(w) + \chi(x^{\mu},w).
\end{eqnarray}
In Eqs. (\ref{P1})-(\ref{P2}), $\delta \Gamma_{AB}^{C}$ and 
$\delta R_{A B}$ are, respectively, the fluctuations of the
Christoffel connections and of the  Ricci tensors, whereas 
$\overline{\Gamma}_{AB}^{C}$ are the  background values of the connections
\begin{equation}
\overline{\Gamma}_{A w}^{C} = {\cal H} 
\delta_{A}^{C},\,\,\,\overline{\Gamma}_{\mu\nu}^{w} = {\cal H}\eta_{\mu\nu},
\end{equation}
where the metric has been assumed in its conformally 
flat parametrization as in Eq. (\ref{m1}).
The metric fluctuation $\delta G_{AB}$ contains scalar, vector and tensor 
modes, namely,
\begin{equation}
\delta G_{AB}(x^{\mu}, w) = \delta G^{(S)}_{AB}(x^{\mu}, w) +\delta 
G^{(V)}_{AB}(x^{\mu}, w) +\delta G^{(T)}_{AB}(x^{\mu}, w).
\end{equation} 
Since the four-dimensional metric is Lorentz invariant, 
the scalar, vector and tensor modes of the higher dimensional geometry
will be classified according to Lorentz transformations as previously 
discussed \cite{t} in a different context \footnote{I am grateful 
to M. Shaposhnikov for valuable comments on this point.}. The simultaneous 
use of gauge-invariance and Lorentz invariance is one of the key points 
of the present analysis.

\subsection{Gauge-invariant variables}

The Bardeen formalism for metric fluctuations \cite{bar} 
will now be generalized 
to the case of higher-dimensional metrics containing non 
compact transverse dimensions of the type of the one given in Eq. (\ref{m1}).
The fluctuation of the metric can  be parametrized as
\begin{equation}
\delta G_{A B}=a^2(w) \left(\matrix{2 h_{\mu\nu} 
+(\partial_{\mu} f_{\nu} +\partial_{\nu} f_{\mu}) 
+ 2\eta_{\mu \nu} \psi
+ 2 \partial_{\mu}\partial_{\nu} E
& D_{\mu} + \partial_{\mu} C &\cr
D_{\mu} + \partial_{\mu} C  & 2 \xi &\cr}\right),
\label{lorf}
\end{equation}
where the Lorentz indices run over the four space-time dimensions.
The tensor $h_{\mu\nu}$ is  traceless and divergence-less 
\begin{equation}
h_{\mu}^{\mu} = 0,\,\,\,\,\,\,\,\,\,\,\,
\partial_{\nu} h_{\mu}^{\nu}= 0,
\end{equation}
 corresponding to five independent components. 
The vectors $D_{\mu}$ and $f_{\mu}$ are both
divergence-less 
\begin{equation}
\partial_{\mu} D^{\mu} =0,\,\,\,\,\,\,\,\,\,\,\,
\partial_{\mu}f^{\mu}=0,
\end{equation}
so that they will have, together, six  independent components. 
The scalars $E$, $\psi$, $C$ and $\xi$ lead to four 
independent components. For infinitesimal 
coordinate transformations 
\begin{equation}
x_{A} \rightarrow \tilde{x}^{A} = x^{A} + \epsilon^{A}, 
\label{shift}
\end{equation}
the fifteen independent metric fluctuations change as 
\begin{equation}
\delta \tilde{G}_{A B} = \delta G_{AB} - \nabla_{A} \epsilon_{B} - \nabla_{B}
\epsilon_{A}, 
\label{liederiv}
\end{equation}
where, for the background metric of Eq. (\ref{m1}), 
$\epsilon_{A} = a^2(w) ( \epsilon_{\mu}, - \epsilon_{w})$. In Eq. 
(\ref{liederiv}) the Lie 
the covariant derivatives are compuetd using the background metric.

The $\epsilon_{\mu}$ can be written as the derivative of a scalar plus a 
vector 
\begin{equation}
\epsilon_{\mu} = \partial_{\mu} \epsilon + \zeta_{\mu},
\label{eps}
\end{equation}
where $\partial_{\mu} \zeta^{\mu} =0$.
 
Using Eqs. (\ref{liederiv}) and (\ref{eps})  the explicit transformation
properties of the perturbed components of the metric for  
infinitesimal coordinate transformations can be found. 
The transverse and traceless
tensors  are gauge-invariant, i.e. they do not change 
for infinitesimal gauge transformations
\begin{equation}
\tilde{h}_{\mu\nu} = h_{\mu\nu},
\label{hl}
\end{equation}
whereas the vector transform as 
\begin{eqnarray}
&& \tilde{f}_{\mu} = f_{\mu} - \zeta_{\mu},
\label{fl}\\
&&\tilde{D}_{\mu} = D_{\mu} - \zeta_{\mu}'.
\label{zeta}
\end{eqnarray}
Finally the scalars will transform as
\begin{eqnarray}
&&\tilde{E} = E - \epsilon,
\label{El}\\
&&\tilde{\psi} = \psi - {\cal H} \epsilon_{w},
\label{psil}\\
&& \tilde{C} = C - \epsilon' + \epsilon_{w},
\label{Cl}\\
&& \tilde{\xi} = \xi + {\cal H} \epsilon_{w} + \epsilon_{w}'.
\label{xil}
\end{eqnarray}
The prime denotes, as usual, derivation with 
respect to the bulk coordinate. 

Gauge-invariant fluctuations corresponding to scalar 
and vector modes of a given geometry can be constructed as it was noticed 
by Bardeen \cite{bar} in a four-dimensional context and later 
generalized to higher (compact) dimensions \cite{mg1,mg2,ab}. 
Since the geometry contains two divergence-less 
vectors and one gauge function $\zeta_{\mu}$, only one 
gauge-invariant variable is allowed.
A possible choice of gauge-invariant vector fluctuation is
\begin{equation}
\tilde{V}_{\mu} = \tilde{D}_{\mu} - \tilde{f}_{\mu}'.
\label{givec}
\end{equation}
Using Eqs. (\ref{fl}) and (\ref{zeta}), it is easy to see that
$\tilde{V}_{\mu} = V_{\mu}$.

Since there are four scalar fluctuations in the metric and 
two gauge functions, i.e. $\epsilon$ and $\epsilon_{w}$, two
gauge invariant variables can be defined.
The gauge-invariant scalar fluctuations can 
be parametrized as
\begin{eqnarray}
&&\tilde{\Psi} = \tilde{\psi} - {\cal H}  ( \tilde{E}' - \tilde{C}), 
\label{giscal0}\\
&& \tilde{\Xi} = \tilde{\xi} - \frac{1}{a} [ a( \tilde{C} - \tilde{E}')]'.
\label{giscal}
\end{eqnarray}
Using Eqs. (\ref{psil}) and (\ref{xil}) together with 
Eqs. (\ref{El}) and (\ref{Cl}) it can be directly 
obtained that $\tilde{\Xi} = \Xi$ and that
$\tilde{\Psi} = \Psi$. 
The bulk scalar field fluctuation transforms under infinitesimal 
coordinate shifts:
\begin{equation}
\tilde{\chi} = \chi - \varphi'\epsilon_{w}.
\end{equation}
The gauge-invariant scalar field fluctuation will be 
\begin{equation}
\tilde{X} = \tilde{\chi} - \varphi' (\tilde{E}' -\tilde{C}).
\label{chigi}
\end{equation}
Eqs. (\ref{givec}) and (\ref{giscal}) are reminiscent 
of the Bardeen potentials \cite{bar}. There are, of course, infinite 
gauge-invariant variables since every combination of gauge-invariant 
variables is also gauge-invariant. The variables defined 
in Eqs. (\ref{givec}) and (\ref{giscal}) have the virtue of obeying
 very simple equations, as it will be shown. 

\subsection{Gauge choices}

It is useful to fix completely the gauge freedom. 
Not every gauge fixes completely the 
coordinate system. In the following, two useful gauge 
choices will be discussed. If the gauge is completely 
fixed, five of the fifteen degrees of freedom appearing 
in the perturbed metric can be eliminated. Once the 
fluctuations are discussed in a given gauge, the 
gauge invariant variables can be computed in that specific gauge.

In order to fix completely the coordinate system, 
 $\epsilon$, $\epsilon_{w}$ and $\zeta_{\mu}$ should be fixed. 
Two gauge choices fixing completely the coordinate 
system will now be discussed.
The first one is  
\begin{equation}
\tilde{E} = 0, ~~~~\tilde{f}_{\mu} =0,~~~~\tilde{\psi} =0.
\label{condod}
\end{equation}
The gauge is called 
off-diagonal since the scalar contribution of the 
off-diagonal entry of the perturbed 
metric, i.e. $\partial_{\mu} C$ is non vanishing.  

Using Eqs. (\ref{condod}), respectively into Eqs. (\ref{fl}), (\ref{El}) and 
(\ref{psil})  the infinitesimal parameters are fixed to be 
\begin{equation}
\epsilon= E, ~~~~~\epsilon_{w} = \frac{\psi}{{\cal H}}, ~~~~~
f_{\mu} = \zeta_{\mu}.
\end{equation}

Another interesting gauge choice is the generalization is the 
longitudinal gauge where all the off-diagonal (scalar) 
components of the perturbed metric are vanishing: 
\begin{equation}
\tilde{E} = 0, ~~~~ \tilde{C} =0, ~~~~\tilde{f}_{\mu} = 0. 
\label{conlon}
\end{equation}
As before, using Eqs. (\ref{conlon}) into eqs. (\ref{El}), (\ref{Cl}) and 
(\ref{fl}) the parameters of the gauge transformation can be 
totally fixed:
\begin{equation}
\epsilon= E, ~~~~~\epsilon_{w} = (E' - C), ~~~~~\zeta_{\mu} = f_{\mu}.
\end{equation}

Not all the gauge choices have the feature of fixing 
completely the coordinate system. As an example, consider the gauge
\begin{equation}
\tilde{\psi} =0,\,\,\,\,\tilde{C} =0, \,\,\,\,\tilde{D}_{\mu} =0.
\label{consin}
\end{equation}
Using the conditions of Eqs. (\ref{consin}) 
into Eqs. (\ref{psil}), (\ref{Cl}) 
and (\ref{zeta}) the parameters of the gauge transformation are 
fixed but only up to arbitrary integration constants depending 
only upon $x_{\nu}$:
\begin{eqnarray}
&&\epsilon = Q(x^{\nu}) + \int \biggl( C + \frac{\psi}{{\cal H}}\biggr)dw,
\nonumber\\
&& \zeta_{\mu} = P_{\mu} ( x^{\nu}) + \int D_{\mu} d w.
\end{eqnarray}
Hence, 
 even after the gauge choice (\ref{consin}) has been imposed, 
$\epsilon$ and $\zeta_{\mu}$ can still be shifted by an arbitrary 
function depending upon $x^{\nu}$. In this sense the gauge-fixing described 
by Eqs. (\ref{consin}) is not complete. Further discussions concerning 
gauge choices in the specific framework of \cite{rs,rs2} are contained in 
\cite{gar}.

\renewcommand{\theequation}{4.\arabic{equation}}
\setcounter{equation}{0}
\section{Evolution equations for the metric fluctuations} 

The evolution equations of the fluctuations will now be derived. 
The coupled system of Eqs. (\ref{P1})--(\ref{P2}) will be diagonalized and 
reduced to a set of (decoupled) second order differential equations. 
Particular attention will be paid to the evolution of the gauge-invariant 
variables whose equations of motion should coincide in different gauges.
This useful feature of the formalism will be used in order 
to cross-check the consistency of the resulting equations which 
will be independently derived in different gauges. 

\subsection{ Perturbations in the longitudinal gauge}

In the longitudinal gauge \footnote{From now on we will drop the 
tilde from the variables unless  strictly required.} the Bardeen 
potentials $\Psi$ and $\Xi$ coincide, respectively, with $\psi$ and $\xi$. 
This simple fact can be easily appreciated by inserting the gauge 
fixing of Eqs. (\ref{conlon}) into Eqs. (\ref{giscal}). Similarly, the
gauge-invariant form of the scalar field fluctuation coincide with $\chi$, 
i.e. $X \equiv \chi$. Finally, from Eq. (\ref{givec}) it can be seen that 
$V_{\mu} \equiv D_{\mu}$ where $V_{\mu}$ is the gauge-invariant 
vector fluctuation defined in Eq. (\ref{givec}).

The fluctuations of the Christoffel connections and of the Ricci tensors 
are reported in Appendix A. Using Eqs. (\ref{Chrlon}) and (\ref{riccilon}) 
into Eqs. (\ref{P1})--(\ref{P2}) the coupled system of evolution equations 
of the metric fluctuations can be obtained: 
\begin{eqnarray}
&& h_{\mu\nu} '' + 3 {\cal H} h_{\mu\nu}' - \partial_{\alpha} 
\partial^{\alpha}h_{\mu\nu} =0,
\label{hII}\\
&&\psi'' + 7 {\cal H} \psi' + {\cal H} \xi' + 
2 ( {\cal H}' + 3 {\cal H}^2) \xi + \frac{1}{3} 
\frac{\partial V}{\partial\varphi} a^2 \chi - \partial_{\alpha}
\partial^{\alpha} \psi =0,
\label{psiII}\\
&& \partial_{\alpha }\partial^{\alpha} D_{\mu} =0,
\label{DII}\\
&& - \partial_{\alpha}\partial^{\alpha} \xi - 4 [ \psi''+ {\cal H} \psi'] - 
4 {\cal H} \xi' - \varphi' \chi' - \frac{1}{3} 
\frac{\partial V}{\partial\varphi} a^2 \chi + \frac{2}{3} V a^2 \xi =0,
\label{xiII}\\
&& \partial_{\alpha} \partial^{\alpha} \chi - \chi'' - 3 {\cal H} \chi' + 
\frac{ \partial^2 V}{\partial\varphi^2} a^2 \chi - 
\varphi' [ 4 \psi' + \xi'] - 2 \xi ( \varphi'' + 3 {\cal H} \varphi') =0.
\label{chiII}
\end{eqnarray}
Eq. (\ref{hII}) can be immediately reduced to a Schr\"odinger-like form 
by defining $ a^{3/2} h_{\mu \nu} = \mu_{\mu\nu}$ 
\begin{equation}
\mu_{\mu\nu}'' - \partial_{\alpha} \partial^{\alpha} \mu_{\mu\nu} - 
\frac{(a^{3/2})''}{a^{3/2}} \mu_{\mu\nu} =0.
\label{sch1}
\end{equation}

Eqs. (\ref{hII})--(\ref{chiII})  
contain second order derivatives with respect to time. 
Eqs. (\ref{hII})--(\ref{xiII}) come from the perturbed form of Einstein 
equations, i.e. Eq. (\ref{P1}). Eq. (\ref{chiII}) is the explicit form of 
Eq. (\ref{P2}).
From the off-diagonal 
components of the Einstein equations a number of constraints, connecting 
the first derivatives of different fluctuations, can be obtained:
\begin{eqnarray}
&&\partial_{\mu}\partial_{\nu}[ \xi - 2 \psi] =0,
\label{con1}\\
&& D_{\mu}' + 3 {\cal H} D_{\mu} =0,
\label{con2}\\
&& 6 {\cal H} \xi + 6 \psi' + \chi \varphi' =0.
\label{con3}
\end{eqnarray}
Eqs. (\ref{con1})--(\ref{con3}) are  crucial in order to 
diagonalize the full system. The constraints (\ref{con1}) and (\ref{con3}) 
pertain to scalar modes. This allows to determine the number of propagating 
degrees of freedom. There are, in this gauge, three scalar functions: 
$\psi$, $\xi$ 
and $\chi$. They obey a set of coupled dynamical 
equations but there are two constraints. 
Hence, there will be only one physical scalar mode.

The equations of motion for the scalar 
modes represented by $ \xi$ , $\psi$ and $\chi$ can be decoupled
using the following procedure.
Eq. (\ref{con1}) 
implies that $\xi = 2 \psi$. By summing up Eq. (\ref{psiII}) 
and Eq. (\ref{xiII}) the following equation can be obtained:
\begin{equation}
\psi'' + {\cal H} \psi' +\partial_{\alpha} \partial^{\alpha} \psi + 
\frac{\chi' \varphi'}{3} =0.
\label{pseq}
\end{equation}
In order to get to  Eq. (\ref{pseq}), the 
background relations (\ref{c1})--(\ref{c2}) 
have been used. 

Using Eq. (\ref{con1}), 
the constraint of Eq. (\ref{con3}) 
can be expressed as 
\begin{equation}
\chi = - \frac{6}{\varphi'} ( \psi' + 2 {\cal H} \psi), 
\end{equation}
which implies, if inserted into Eq. (\ref{pseq}), that 
\begin{equation}
\psi'' + [3 {\cal H} - 2 \frac{\varphi''}{\varphi'}] \psi' + 
[ 4 {\cal H}' - 4 {\cal H} \frac{\varphi''}{\varphi'}] \psi  - 
\partial_{\alpha}\partial^{\alpha} \psi=0.
\label{pseq2}
\end{equation}
Since, in the longitudinal gauge, $\Psi= \psi$ 
Eq. (\ref{pseq2}) holds also for $\Psi$.

Defining the rescaled variable 
\begin{equation}
\Phi = \frac{a^{3/2}}{\varphi'} \Psi,
\end{equation}
from Eq. (\ref{pseq2}) the following equation can be 
obtained (see Appendix B):
\begin{equation}
\Phi'' - \partial_{\alpha} \partial^{\alpha} \Phi - z 
\biggl( \frac{1}{z}\biggr)'' \Phi =0,
\label{phieq}
\end{equation}
where 
\begin{equation}
z = \frac{a^{3/2} \varphi'}{{\cal H}},
\end{equation}
is an interesting function which controls the localization 
properties of the scalar zero modes as it will be discussed in 
Section V.
An equation analogous to (\ref{pseq2}) holds also for the other 
Bardeen potential, namely $\Xi$, thanks to the constraint of Eq. (\ref{con1}).

Inserting Eq. (\ref{con1}) and recalling Eqs. (\ref{c1})--(\ref{c2}) also 
the equation for $\chi$ can be expressed in a more tractable form , namely:
\begin{equation}
\partial_{\alpha} \partial^{\alpha} \chi - \chi'' - 3 {\cal H} \chi 
+ \frac{\partial^2 V}{\partial\varphi^2} a^2 \chi  - 6 \varphi' \psi' - 
4 \psi \frac{\partial V}{\partial\varphi} a^2 =0.
\label{chi1}
\end{equation}

Eq. (\ref{chi1})  can be reduced to a Schr\"odinger-like form 
in terms of the following variable:
\begin{equation}
{\cal G} = a^{3/2} \chi - z \psi, ~~~~~~z = \frac{a^{3/2} \varphi' }{{\cal H}}
\label{canonical}
\end{equation}
The equation obeyed by ${\cal G}$ is simply :
\begin{equation}
{\cal G}'' - \partial_{\alpha} \partial^{\alpha} {\cal G} 
- \frac{z''}{z} {\cal G} =0.
\label{caneq}
\end{equation}
Eq. (\ref{caneq}) can be obtained by expressing 
$\psi$ as 
\begin{equation}
\psi =  \frac{a^{3/2}}{z} \chi -\frac{{\cal G}}{z}.
\end{equation}
From this last equation, 
the derivatives of $\psi$ with respect to $w$ 
\begin{equation}
\psi'(\chi, \chi', {\cal G}, {\cal G}'), \,\,\,\,
\psi''(\chi, \chi',\chi'', {\cal G}, {\cal G}', {\cal G}''),
\end{equation}
can be inserted back into Eq. 
(\ref{pseq}). 
Using now 
simultaneously the constraints (\ref{con3}) and Eq. (\ref{chi1}), 
 Eq. (\ref{caneq}) is obtained
after some algebra involving the repeated use of 
Eqs. (\ref{c1})--(\ref{c2}).

The canonical variable ${\cal G}$ is gauge-invariant. Recall 
in fact that under infinitesimal gauge transformations 
\begin{eqnarray}
&&\tilde{\chi} = \chi - \epsilon_{w} \varphi',
\nonumber\\
&& \tilde{\psi} = \psi - \epsilon_{w} {\cal H},
\end{eqnarray}
so that $\tilde{{\cal G}} = {\cal G}$. 
Recall now that the gauge-invariant fluctuations  connected with 
$\psi$ and $\chi$ (i.e. $\Psi$ and $X$) 
are given, respectively, by Eqs. (\ref{giscal0}) and 
(\ref{chigi}). Since it has been shown that ${\cal G}$ is gauge-invariant, 
it is possible to express it as a direct combination of gauge-invariant 
quantities
\begin{equation}
{\cal G} = a^{3/2} X - z \Psi.
\label{giG}
\end{equation}

\subsection{Perturbations in the off-diagonal gauge}

The evolution equations for the coupled system 
given in Eqs. (\ref{P1})--(\ref{P2}) will now be studied in the off-diagonal 
gauge. The interest of this exercise is not only academic since it represents
an important cross-check of the validity of the formalism. In fact, by 
definition of gauge-invariant variable, 
the evolution of the Bardeen potentials should have the same form 
in any gauge.

The fluctuations of the Christoffel connections and of the Ricci tensors are 
reported in the Appendix A.
Using Eqs. (\ref{Chrod}) and (\ref{ricciod}) into 
Eqs. (\ref{P1})--(\ref{P2}) the 
explicit form of the perturbed 
equations of motion in the off-diagonal gauge reads:
\begin{eqnarray}
&& h_{\mu\nu} '' + 3 {\cal H} h'_{\mu\nu} - \partial_{\beta}\partial^{\beta} 
h_{\mu \nu} =0,
\label{hod}\\
&& {\cal H} \xi' + 2 ( {\cal H}' + 3 {\cal H}^2)\xi - {\cal H} 
\partial_{\beta}\partial^{\beta} C 
+ \frac{1}{3} \frac{\partial V}{\partial\varphi} a^2 \chi =0,
\label{xiod}\\
&& \partial_{\beta} \partial^{\beta} D_{\mu} =0,
\label{Dod2}\\
&& - \partial_{\alpha} \partial^{\alpha} \xi - 4 {\cal H} \xi' + {\cal H} 
\partial_{\alpha} \partial^{\alpha} C + (\partial_{\alpha } 
\partial^{\alpha}C)' - \varphi' \chi' - \frac{1}{3} 
\frac{\partial V}{\partial\varphi} a^2 \chi + \frac{2}{3} V a^2 \xi =0,
\label{Cod}\\
&& \partial_{\alpha} \partial^{\alpha} \chi - \chi'' - 3 {\cal H} \chi ' + 
\frac{\partial^2 V}{\partial\varphi^2} a^2 \chi + \varphi' ( \partial_{\alpha} 
\partial^{\alpha} C - \xi') - 2 \xi ( \varphi'' + 3 {\cal H} \varphi') =0.
\label{chiod}
\end{eqnarray}
Eqs. (\ref{hod})--(\ref{Cod}) are derived from Eq. (\ref{P1}), whereas, Eq. 
(\ref{chiod}) is derived from Eq. (\ref{P2}) and it is the perturbed form of 
the scalar field equation.
The constraints connecting the first derivatives of the fluctuations will be:
\begin{eqnarray}
&& {\cal \xi} - 3 {\cal H} C - C'=0,
\label{xiconod}\\
&& D_{\mu}' + 3 {\cal H} D_{\mu} =0,
\label{Dod}\\
&& 6 {\cal H} \xi + \chi \varphi' =0.
\label{conod}
\end{eqnarray}
Summing up Eqs. (\ref{xiod}) and (\ref{Cod}) the following equation 
is obtained:
\begin{equation}
{\cal H}\xi' - \partial_{\alpha} \partial^{\alpha}\xi - 4 {\cal H} \xi' + 
(\partial_{\alpha} \partial^{\alpha})' - \varphi' \chi' =0.
\label{step1}
\end{equation}
Using Eq. (\ref{xiconod}), the constraint given in Eq. 
 (\ref{conod}) can be expressed as 
\begin{equation}
\chi = - \frac{6 {\cal H}}{\varphi'} ( 3 {\cal H} C + C').
\label{step2}
\end{equation}
Inserting Eq. (\ref{step2}) into Eq. (\ref{step1}) 
\begin{equation}
C'' + 3 {\cal H} C' + 3 {\cal H}' C + \partial_{\alpha}\partial^{\alpha} 
C + \frac{ \varphi'\chi'}{3 {\cal H}} =0.
\end{equation}
Finally, using  the constraint provided by Eq. (\ref{step2}) in order to 
eliminate $\chi$  the following decoupled equation can be obtained:
\begin{equation}
C'' + \biggl[ 3 {\cal H} + 2 \frac{{\cal H}'}{\cal H} - 2 
\frac{\varphi''}{\varphi'}\biggr]C' + 3\biggl[3 {\cal H}' 
 - 2{\cal H}\frac{\varphi''}{\varphi'}\biggr] C - 
\partial_{\alpha} \partial^{\alpha} C =0.
\label{cdec}
\end{equation}
 
In order to  check for the consistency of the formalism  the 
evolution equation for the gauge-invariant variables can be obtained. 
In the 
off-diagonal gauge $\Psi = {\cal H} C$. From Eq. (\ref{cdec}) the equation 
for $\Psi$ is derived after the repeated use of the background relations given 
in Eqs. (\ref{c1}) and (\ref{c2}):
\begin{equation}
\Psi'' + [3 {\cal H} - 2 \frac{\varphi''}{\varphi'}] \Psi' + 
[ 4 {\cal H}' - 4 {\cal H} \frac{\varphi''}{\varphi'}] \Psi  - 
\partial_{\alpha}\partial^{\alpha} \Psi=0.
\label{cdec2}
\end{equation}
Notice, as it should, that this equation is exactly Eq. (\ref{pseq2}) 
obtained in the longitudinal gauge. This shows that the equations of the 
gauge-invariant variables are the same in any gauge {\em provided} the 
gauge freedom is completely fixed in each of the selected gauges.

Finally, in the off-diagonal gauge, the gauge-invariant 
normal mode ${\cal G}$ defined in Eq. (\ref{giG}) becomes
\begin{equation}
{\cal G} = a^{3/2} \chi.
\end{equation}
Hence, gauge-invariance suggests that also in the 
 off-diagonal gauge the equation for ${\cal G}$ will be  given 
by Eq. (\ref{caneq}). In fact,  the constraints 
(\ref{xiconod}) and (\ref{conod}) can be used into Eq. (\ref{chiod}) 
in order to eliminate the dependence on $\xi$ and $C$. The resulting 
equation for $a^{3/2} \chi$ is exactly Eq. (\ref{caneq}) once we recall that, 
from Eq. (\ref{s}), the double derivative of the potential 
with respect to $\varphi$ can be expressed as 
\begin{equation}
a^2 \frac{\partial^2 V}{\partial\varphi^2} = \biggl[ {\cal H} 
\frac{\varphi''}{\varphi'} - 6 {\cal H}^2 + \frac{ \varphi'''}{\varphi'} + 
3 {\cal H}' \biggr].
\end{equation}

\subsection{Canonical normal modes of the action}

The investigation of the canonical normal modes of the 
action (\ref{totaction}) is rather long and, here, only the 
 results will 
be reported. The same relations obtained from the 
equations of motion could be obtained by perturbing the 
action (\ref{totaction}) 
to second order in the amplitude of the fluctuations.
In discussing the fluctuations of   (\ref{totaction}) 
it is better to perturb
 the Einstein-Hilbert term in the form
\begin{equation}
G^{A B}\biggl( \Gamma_{A C}^{D} \Gamma_{B D}^{C}
- \Gamma_{A B}^{C} \Gamma_{C D}^{D}\biggr) 
\end{equation}
to second order in the amplitude of the fluctuations. This form of the 
gravity action automatically eliminates the total derivatives.

The normal modes for tensors turn out to be for each of the 
polarizations
\begin{equation}
\delta^{(2)} S_{T} = \int d^4 x d w \frac{1}{2}\biggl[ 
\eta^{\alpha\beta} 
\partial_{\alpha} \mu \partial_{\beta} \mu -{\mu'}^2
- \frac{(a^{3/2})''}{a^{3/2}} \mu^2\biggr]
\label{canten}
\end{equation}
where $\mu = a^{3/2} h$ and $\eta_{\alpha\beta}$ is the Minkowski metric.

Fixing the longitudinal gauge
the normal modes for the vector action will be 
\begin{equation}
\delta^{(2)} S_{V} = \int d^4 x d w \frac{1}{2}\biggl[ \eta^{\alpha\beta} 
\partial_{\alpha} {\cal D}^{\mu} \partial_{\beta} {\cal D}_{\mu} \biggr]
\label{canvec}
\end{equation}
where ${\cal D}_{\mu} = a^{3/2} D_{\mu}$.
The normal modes for the scalar action will be 
\begin{equation}
\delta^{(2)} S_{T} = \int d^4 x d w \frac{1}{2}\biggl[ 
\eta^{\alpha\beta} 
\partial_{\alpha} {\cal G} \partial_{\beta}{\cal G} - {{\cal G}'}^2
- \frac{z''}{z} {\cal G}^2\biggr]
\label{canscal}
\end{equation}
where ${\cal G}$ is the variable previously defined and 
$z = a^{3/2} \varphi' /{\cal H}$. In order to get these normal modes the 
longitudinal gauge has been selected. The calculation can be also done 
without assuming any specific gauge (as discussed in \cite{mg1}, in a 
different context). The advantage of working with the full (i.e. non gauge
-fixed) action is that the constraints will naturally appear.

Suppose, for instance, that the  action
 of Eq. (\ref{totaction})
is perturbed to second order in the amplitude of the vector fluctuations
without  fixing the specific gauge. Hence, in the perturbed action, after 
integration by parts and up to total derivative terms, a term like 
\begin{equation}
\int a^3 d^4 x ~ d w ~~ \partial_{(\mu}f_{\nu)} \biggl[
 {\partial^{(\mu}D^{\nu)}}' + 3 {\cal H}  \partial^{(\mu}D^{\nu)}\biggr]
\end{equation}
will appear on top of the kinetic term for $D_{\mu}$. By taking the 
derivative of the perturbed action with respect to $\partial_{\mu}f_{\nu}$ 
the constraint of Eq. (\ref{con2}) can be obtained.

\renewcommand{\theequation}{5.\arabic{equation}}
\setcounter{equation}{0}
\section{Zero modes of the fluctuations, localization and normalizability} 

\subsection{ Localization of the tensor modes}
The evolution equation of each tensor polarization can be written as 
\begin{equation}
\mu'' - \partial_{\alpha}\partial^{\alpha} \mu - \frac{s''}{s}\mu=0.
\label{tenseq}
\end{equation}
where $s h = \mu$ and with $s = a^{3/2}$.
The zero mode of this equation corresponds to 
\begin{equation}
\mu_0(w) = {\rm K} a^{3/2},
\end{equation}
where $K$ is a constant which should be fixed by normalization.
The normalization condition reads 
\begin{equation}
\int |\mu_0(w)|^2 dw ={\rm K}^2 \int_{-\infty}^{+\infty} a^3(w) dw =1. 
\end{equation}
The zero mode is then normalized provided $a^3(w)$ goes to zero to infinity 
faster than $1/w$
\begin{equation}
\lim_{|w|\rightarrow  \infty} a^3(w) = {\cal O} \biggl( \frac{1}{w^{\lambda}} 
\biggr),\,\,\,\,\,\, \lambda > 1,
\label{relten}
\end{equation} 
and provided $a(w)$ is regular in all the domain of definition of $w$.
Suppose, for, instance, that $a(w) = (b^2 w^2 + 1)^{-1/2}$ as obtained, 
in a particular case, in Eq. (\ref{s1}). In this 
case ${\rm K} = \sqrt{2 b}$.

The massive modes of Eq. (\ref{tenseq}) obey a Schr\"odinger-like equation 
which can be written as 
\begin{equation}
- \frac{d^2\mu_{m}}{d w^2} + V(w) \mu_{m} = m^2 \mu_{m}, 
\label{sten}
\end{equation}
where the effective ``potential'' term is
\begin{equation}
V(w) = \frac{s''}{s} \equiv {\cal J}^2 - {\cal J}', \,\,\,\,\,{\cal J} = 
-\frac{s'}{s}
\label{pott}
\end{equation}
In Eq. (\ref{pott}) ${\cal J}$ is the superpotential usually defined in the 
context of supersymmetric quantum mechanics \cite{sqm}. 
Using supersymmetric 
quantum mechanics Eq. (\ref{sten}) can be written as
\begin{equation}
{\cal A}^{\dagger} {\cal A} \mu_{m} = m^2 \mu_{m},
\label{susq}
\end{equation}
where
\begin{equation}
{\cal A}^{\dagger} = \biggl( - \frac{d}{dw} + {\cal J} \biggr), \,\,\,\,\,\,
{\cal A} = \biggl( \frac{d}{dw} + {\cal J} \biggr).
\label{operators}
\end{equation}
The fact that the Schr\"odinger equation can be written in this form excludes
the presence of tachyonic modes  with $m^2 <0$. 

\subsection{Localization of the vector zero mode}

The equation describing the vector modes are of the type 
\begin{equation}
\partial_{\alpha}\partial^{\alpha} D =0, \,\,\,\, D' + 3 {\cal H} D=0,
\label{Veq}
\end{equation}
for each polarization of the gauge-invariant vectors. 
Since the equations are first order, the normal modes can be 
read-off from the perturbed action of Eq. (\ref{canvec}). They are 
$ {\cal D} = a^{3/2} D$. Since, from Eq. (\ref{Veq}) 
$D_0 = {\rm K} /a^{3}$ the 
normalization relation for the zero mode will be 
\begin{equation}
\int d w |{\cal D}_0(w)|^2 = {\rm K}^2\int_{\infty}^{+\infty} 
\frac{d w}{a^3(w)}= 1. 
\end{equation}
This relation would imply that 
\begin{equation}
\lim_{|w|\rightarrow  \infty} \frac{1}{a^3(w)} 
= {\cal O} \biggl( \frac{1}{w^{\lambda}} 
\biggr),\,\,\,\,\,\, \lambda > 1.
\label{relvec}
\end{equation} 
Eq. (\ref{relvec}) cannot be true together with Eq. (\ref{relten}). 
Eq. (\ref{relten}) has to be correct for independent reasons: it guarantees
that the four-dimensional Planck mass is finite since, in our parametrization
\footnote{Notice that we always worked in natural five-dimensional units where 
$2\kappa = 16\pi/M_{5}^3 = 1$.}
\begin{equation}
M_{P}^2 \sim M_{5}^3 \int a^3(w) dw.
\end{equation}
Hence, the zero mode of the vector fluctuations is not 
localized on the brane if the four-dimensional Planck mass is finite.

\subsection{Localization of the scalar zero  modes}
The evolution equation for the Bardeen potentials can be 
written as 
\begin{equation}
\Phi'' - \partial_{\alpha}\partial^{\alpha} \Phi - 
z \biggl( \frac{1}{z}\biggr)'' \Phi=0.
\label{Pheq}
\end{equation}
The same equation holds also for the rescaled $\Xi$.
From Eq. (\ref{Pheq}) the zero mode is 
given by $\Phi_0 = {\rm K}/z(w)$ where K is a constant. 
The normalization condition 
is given by, 
\begin{equation}
\int |\Phi_0(w)^2| dw = {\rm K}^2 \int \frac{d w}{z^2} \equiv {\rm K}^2
\int_{-\infty}^{+\infty} \frac{{\cal H}^2}{a^3\,{\varphi'}^2} =1,
\label{normsca}
\end{equation}
which implies that the scalar 
zero mode is normalizable provided $1/z^2(w)$ goes to zero for $|w| 
\rightarrow\infty$ faster than $1/w$, i.e. 
\begin{equation}
\lim_{|w| \rightarrow \infty} \frac{{\cal H}^2}{a^3 {\varphi'}^2} 
= {\cal O}\biggl(\frac{1}{w^{\lambda}}\biggr),\,\,\,\,\lambda > 1,
\end{equation}
and provided $z(w)$ is everywhere regular.
In the background given by Eqs. (\ref{s1})--(\ref{s1a}) the 
scalar zero mode is not normalizable. In fact the normalizability condition 
requires that 
\begin{equation}
\frac{{\rm K}^2}{6} \int_{-\infty}^{+\infty}   
d w\,  b^2\, w^2\, (1 + b^2 w^2)^{3/2},
\end{equation}
is finite. This does not happen since the integrand is not convergent for 
$ |w|\rightarrow \infty$. Since Eq. (\ref{Pheq}) is second order there will 
be also a second (linearly independent) solution to Eq. (\ref{Pheq}) for the 
zero mode, namely $\Phi_{0}(w) \sim z^{-1}(w) \int^{w} z^2(w') dw'$. If the 
first solution (going as $z^{-1}(w)$) is not normalizable, also the second 
one will not be normalizable \cite{mp}. For instance, in the case 
of the solution (\ref{s1})--(\ref{s1b})
 the second zero mode blows up, for large $|w|$ 
as $|w|^{5/2}$.

The massive modes of $\Phi$ will obey a Schr\"odinger-like equation 
\begin{equation}
- \frac{d^2 \Phi_{m}}{dw^2} + V_{\Phi}(w) \Phi_{m} = m^2 \Phi_{m}, 
\label{sscal}
\end{equation}
whose associated effective potentials and superpotentials are 
\begin{equation}
V_{\Phi}(w)= {\cal J}^2 - {\cal J}' ,\,\,\,\, 
J_{\Phi}(w) = \frac{z'}{z}.
\end{equation}

The equation for the canonical normal mode ${\cal G}$ reads
\begin{equation}
{\cal G}'' - \partial_{\alpha}\partial^{\alpha} {\cal G} - 
\frac{z''}{z} {\cal G}=0.
\label{Geq}
\end{equation}
which implies that the zero mode should go as 
${\cal G}(w) = {\rm K} z(w)$. This implies that 
the normalization condition is
\begin{equation}
{\rm K}^2 \int dw |z(w)|^2 = {\rm K}^2\int_{-\infty}^{+\infty} 
\frac{a^3{ \varphi'}^2}{{\cal H}^2} =1
\label{normnorm}
\end{equation}
This normalization integral implies that 
$z(w)$ has to be always finite for any $w$ and that 
it should go to zero, for large $|w|$, faster than $1/w$.
Since the integrand appearing in Eq. (\ref{normnorm}) 
is exactly the inverse of the one appearing in Eq. (\ref{normsca}), 
it is clear that they cannot be simultaneously finite for a given physical
model. However, they can be simultaneously divergent. An example is, again 
the model discussed in Eqs. (\ref{s1})--(\ref{s1a}). In this case the 
integrand appearing in Eq. (\ref{normnorm}) is, up to numerical factors,
\begin{equation}
\frac{1}{b^2 w^2 ( b^2 w^2 + 1)^2},
\end{equation}
which is not convergent for $|w| \rightarrow 0$.

The massive modes related to ${\cal G}$ can be, again, described 
by a Schr\"odinger-like equation of the type of the one reported in Eq. 
(\ref{sscal}). Notice that ther is an interesting ``duality'' relation
among the effective potentials. The effective potential for the equation for 
${\cal G}$ is simply
\begin{equation}
V_{{\cal G}}= \frac{z''}{z}.
\end{equation}
For $z(w)\rightarrow 1/z(w)$,
\begin{equation}
V_{{\cal G}}(w) \rightarrow V_{\Phi}(w) = 
z(z^{-1})''.
\end{equation}
In terms of the superpotential this symmetry implies 
that for $z \rightarrow 1/z$
\begin{equation}
V_{\Phi}(w) = {\cal J} - {\cal J}' \,\,\,
\rightarrow\,\,\, V_{{\cal G}}(w) = {\cal J}^2 +  {\cal J}'.
\end{equation}
The two Schr\"odinger-like equations obeyed by the massive 
modes of ${\cal G}$ and $\Phi$ can then be written, in terms of the operators 
defined in Eq. (\ref{operators}) as 
\begin{eqnarray}
&& {\cal A}^{\dagger} {\cal A} \Phi_{m} = m^2_{\Phi} \Phi_{m},
\nonumber\\
&&   {\cal A} {\cal A}^{\dagger} {\cal G}_{m} = m^2_{\cal G} {\cal G}_{m},
\end{eqnarray}
with ${\cal J}= z'/z$. In the context of 
supersymmetric quantum mechanics these two potentials $V_{\Phi}$ and 
$V_{\cal G}$ are supersymmetric  partners potentials \cite{sqm}.
This property implies that the spectra are related.

\renewcommand{\theequation}{6.\arabic{equation}}
\setcounter{equation}{0}
\section{Concluding remarks} 

If physical walls in five dimensions are described 
using Eqs. (\ref{s})--(\ref{e2}), then in order 
to have localized zero modes of the metric the following 
integrals should be simultaneously convergent:
\begin{eqnarray}
&& I_{{\rm tens}}= \int_{-\infty}^{+\infty} a^3(w) dw,
\label{first1}\\
&& I_{\rm vec} =
\int_{-\infty}^{+\infty} \frac{d w}{a^3(w)} ,
\label{first2}\\
&& I_{\Phi}= \int_{-\infty}^{+\infty} \frac{dw}{z^2(w)},
\label{sec1}\\ 
&& I_{{\cal G}}=\int_{-\infty}^{+\infty} z^{2}(w) dw,
\label{sec2}
\end{eqnarray}
where 
\begin{equation}
z(w) = \frac{a^{3/2} \varphi'}{{\cal H}},
\end{equation}
These equations do not assume any specific background solution
but only the form (\ref{m1}) of the metric 
together with Eqs. (\ref{s})--(\ref{e2}). 

The integrals of Eqs. (\ref{first1})--(\ref{sec2}) cannot 
be all simultaneously convergent. Since the 
four dimensional Planck mass should be finite, the integral in 
Eq. (\ref{first1}), i.e. $I_{\rm tens}$, should converge. This implies that 
tensor modes are localized on the brane and that they lead to 
ordinary four-dimensional gravity. The convergence of 
the integral in Eq. (\ref{first2}), i.e. $I_{\rm vec}$  would insure 
the normalizability of the vector zero mode. This is not 
possible as long as the four dimensional Planck mass is finite.

Using the explicit form of $z(w)$, and recalling that, from 
Eq. (\ref{c2}),  $\varphi'^2 = 6 ({\cal H}^2 - {\cal H}')$, 
Eq. (\ref{sec1}) can be written as 
\begin{equation}
I_{\Phi} =  
 \frac{1}{6} \int_{-\infty}^{+\infty} 
\frac{{\cal H}^2}{ a^3({\cal H}^2 - {\cal H}')} dw.
\label{sccon}
\end{equation}
From Eq. (\ref{first1}) $a(w)$, should decay at infinity. If 
$a(w) \sim w^{- \alpha}$, for $|w|\rightarrow \infty$,
 ${\cal H}^2/({\cal H}^2 - {\cal H}')$ goes to a constant 
and the integrand blows up as $w^{3\alpha}$. If $a(w)$ decays exponentially 
the integrand 
appearing in (\ref{sccon}) blows up exponentially.

Eq. (\ref{sec2}) can be written as   
\begin{equation}
I_{\cal G}= \int_{-\infty}^{+\infty} \frac{a^3 \varphi'^2}{{\cal H}^2} 
dw \equiv
6\int_{-\infty}^{+\infty} \biggl\{a^3 
\frac{ {\cal H}^2 - {\cal H}'}{{\cal H}^2}\biggr\}.
\end{equation}
If $I_{\rm tens}$ is convergent at infinity, 
then $I_{\cal G}$ will be convergent in the same limit but this 
does not imply the localization as discussed in Section V.

The integral relations obtained in the present 
investigation are of general relevance for warped scalar-tensor backgrounds
and they can be exploited in order to check for the normalizability 
of the scalar and vector zero modes of the geometry.  
Depending upon the specific model, the range of variation of 
$w$ can be changed. However, the localization of the zero modes 
will always be conrtoled by the functions derived in the present analysis.

In this paper, the Bardeen formalism has been 
extended to the case of warped metrics and the localization 
of scalar and vector modes of the geometry has been investigated. 
All the results 
have been expressed in terms of gauge-invariant, i.e. coordinate independent, 
variables. The coupled system of gauge-invariant perturbations has
been reduced to a set of Schr\"odinger-like equations whose 
effective potential determines the 
localization properties of the corresponding zero modes. 
Supersymmetric quantum mechanics has been used in order to 
show the absence of tachyonic states in the spectrum and duality 
relations among the effective potentials have been derived.  Possible 
extensions of the present results to related problems are in progress.

The results of the present investigation can be generalized to higher 
dimensions. Of particular interest, in this framework, are six-dimensional 
models which are relevant both in the context of compact \cite{ced1,ced2,ced3} 
and non compact \cite{nceda,nced1,nced1a,nced2,nced2a,mmm} 
extra-dimensions. Six-dimensional 
(non-singular) brane solutions have been recently derived in the 
context of the Abelian-Higgs model \cite{gms}  and their fluctuations 
can be investigated with suitable extensions of the techniques discussed 
in the present analysis.
 
\section*{Acknowledgments}
The author is deeply indebted to M. E. Shaposhnikov for inspiring discussions 
and valuable comments. The author wishes also to thank G. Veneziano and M. 
Gasperini for their kind interest in the subject of the present analysis.

\newpage
\begin{appendix}

\renewcommand{\theequation}{A.\arabic{equation}}
\setcounter{equation}{0}
\section{Fluctuations in the longitudinal and off-diagonal gauges} 
In the longitudinal gauge the fluctuation of the metric is:
\begin{equation}
\delta G_{A B}=a^2(w) \left(\matrix{2 h_{\mu\nu} + 2 \eta_{\mu\nu} \psi
& D_{\mu} &\cr
D_{\mu}   & 2 \xi &\cr}\right).  
\label{longm1}
\end{equation}
The Christoffel connections perturbed to first order in $\delta G_{A B}$ are 
\begin{eqnarray}
&&\delta^{(1)} \Gamma_{w w}^{w} = - \xi' ,
\nonumber\\
&&\delta^{(1)} \Gamma_{\mu w}^{w} = - \partial_{\mu} \xi + {\cal H} D_{\mu},
\nonumber\\
&& \delta^{(1)} \Gamma_{\mu\nu}^{w} = h_{\mu\nu}' + 2 {\cal H} h_{\mu\nu} 
+ \eta_{\mu\nu} [ \psi' + 2 {\cal H} \psi + 2 {\cal H} \xi]
- \frac{1}{2} (\partial_{\mu} D_{\nu} + \partial_{\nu} D_{\mu}),
\nonumber\\
&& \delta^{(1)} \Gamma_{\alpha\beta}^{\mu} =  - \partial^{\mu} h_{\alpha\beta}
+ \partial_{\alpha} h_{\beta}^{\mu} + \partial_{\beta} h_{\alpha}^{\mu} 
+ \delta_{\beta}^{\mu} \partial_{\alpha} \psi + \delta_{\alpha}^{\mu} 
\partial_{\beta} \psi + \eta_{\alpha\beta} [ - {\cal H} D^{\mu} - 
\partial^{\mu} \psi],
\nonumber\\
&& \delta^{(1)}\Gamma_{ww}^{\mu} = - \partial^{\mu} \xi
 + {D^{\mu}}' + {\cal H} 
D^{\mu},
\nonumber\\
&& \delta^{(1)}\Gamma_{w\alpha}^{\mu} = {h_{\alpha}^{\mu}}' 
+ \delta_{\alpha}^{\mu} 
\psi' + \frac{1}{2} (\partial_{\alpha} D^{\mu} - \partial^{\mu} D_{\alpha}).
\label{Chrlon}
\end{eqnarray}
Using Eqs. (\ref{Chrlon}) the Ricci tensors perturbed to first order 
in the metric fluctuations can be computed with some trivial algebra 
involving the use of the Palatini identities:
\begin{eqnarray}
\delta^{(1)} R_{\mu\nu} &=& h_{\mu\nu}'' + 3 {\cal H} h_{\mu\nu}' + 
h_{\mu\nu} ( 2 {\cal H}' + 6 {\cal H}^2) - 
\partial_{\alpha}\partial^{\alpha} h_{\mu\nu} 
+ \partial_{\mu}\partial_{\nu} [ \xi - 2 \psi] 
\nonumber\\
&+& 
\eta_{\mu\nu} [ \psi'' + 7 {\cal H} \psi' + 
( 2 {\cal H}' + 6 {\cal H}^2)(\xi + \psi) - 
\partial_{\alpha} \partial^{\alpha} \psi + {\cal H} \xi' ]
\nonumber\\
&-& \frac{1}{2} [( \partial_{\mu} D_{\nu} + \partial_{\nu} D_{\mu} )' + 
3 {\cal H} ( \partial_{\mu} D_{\nu} + \partial_{\nu} D_{\mu})],
\nonumber\\
\delta^{(1)} R_{\mu w} &=& \partial_{\mu} [ - 3 {\cal H} \xi - 3 \psi' ] 
- \frac{1}{2} \partial_{\alpha} \partial^{\alpha} D_{\mu}  +
D_{\mu} ( {\cal H}' + 3 {\cal H}^2),
\nonumber\\
\delta^{(1)} R_{w w} &=& - \partial_{\alpha} \partial^{\alpha} \xi - 
4 ( \psi'' + {\cal H} \psi') - 4 {\cal H} \xi',
\label{riccilon}
\end{eqnarray}
where the background value of the Christoffel connections
\begin{equation}
\overline{\Gamma}^{A}_{B w}= {\cal H} \delta_{A}^{B}, \,\,\,\,\,\, 
\overline{\Gamma}^{w}_{\alpha\beta}
={\cal H} \eta_{\alpha\beta}. 
\end{equation}
has been used.

In the off-diagonal gauge the perturbed metric takes the form
\begin{equation}
\delta G_{A B}=a^2(w) \left(\matrix{2 h_{\mu\nu} 
& D_{\mu} + \partial_{\mu} C &\cr
D_{\mu} + \partial_{\mu} C  & 2 \xi &\cr}\right).  
\label{lorf2}
\end{equation}
Following the same steps outlined in the 
discussion of the longitudinal gauge, 
 the fluctuation of the Christoffel connection can 
be easily  obtained
\begin{eqnarray}
&& \delta^{(1)} \Gamma_{w w}^{w} = - \xi',
\nonumber\\
&&\delta^{(1)} \Gamma_{\mu \nu}^{w} = h_{\mu\nu}' + 2 {\cal H} h_{\mu\nu} - 
\partial_{\mu}\partial_{\nu} C - \frac{1}{2} \biggl( \partial_{\mu} D_{\nu} 
+ \partial_{\nu} D_{\mu} \biggr) + 2 \eta_{\mu\nu} {\cal H} \xi,
\nonumber\\
&& \delta^{(1)} \Gamma_{\mu w}^{w} = - \partial_{\mu} \xi + {\cal H} 
( \partial_{\mu} C + D_{\mu}),
\nonumber\\
&& \delta^{(1)} \Gamma_{\alpha\beta}^{\mu} = \biggl( 
- \partial^{\mu} h_{\alpha\beta}
+ \partial_{\beta} h^{\mu}_{\alpha} + \partial_{\alpha} h_{\beta}^{\mu} \biggr)
- {\cal H} \eta_{\alpha\beta} ( \partial^{\mu} C + D^{\mu}),
\nonumber\\
&& \delta^{(1)} \Gamma_{w w}^{\mu} = -\partial^{\mu} \xi + 
{\cal H} (\partial^{\mu} 
C + D^{\mu}) + (\partial^{\mu} C + D^{\mu} )' ,
\nonumber\\
&& \delta^{(1)} \Gamma_{w \alpha }^{\mu} = {h_{\alpha}^{\mu}}' + \frac{1}{2} ( 
\partial_{\alpha} D^{\mu} - \partial^{\mu} D_{\alpha} ).
\label{Chrod}
\end{eqnarray}

From Eq. (\ref{Chrod}) the fluctuations 
of the Ricci tensors are, after some algebra,  
\begin{eqnarray}
\delta^{(1)} R_{\mu\nu} &=& [h_{\mu\nu}'' + 3 {\cal H} h_{\mu\nu}' 
+  h_{\mu\nu} ( 2 {\cal H}' + 6 {\cal H}^2) 
\partial_{\beta}\partial^{\beta} h_{\mu\nu}]  + \eta_{\mu\nu} [ {\cal H} \xi'
 + 2 ({\cal H}' + 3 {\cal H}^2) \xi 
- {\cal H} \partial_{\beta}\partial^{\beta} C]
\nonumber\\
&+& \partial_{\mu}\partial_{\nu} [ \xi - 3 {\cal H} C - C']
-\frac{1}{2} [ ( \partial_{\mu} D_{\nu} + \partial_{\nu} D_{\mu})' + 
3 {\cal H} (\partial_{\nu} D_{\mu} + \partial_{\mu} D_{\nu} )],
\nonumber\\
\delta^{(1)} R_{ww} &=& 
- \partial_{\alpha}\partial^{\alpha} \xi - 4 {\cal H} \xi' +
{\cal H} \partial_{\alpha} \partial^{\alpha} C 
+ (\partial_{\alpha} \partial^{\alpha}C)',
\nonumber\\
\delta^{(1)} R_{\mu w} &=& - \frac{1}{2} 
\partial_{\alpha}\partial^{\alpha} D_{\mu} 
+ ({\cal H}' + 3 {\cal H}^2) D_{\mu} + 
\partial_{\mu} [ - 3 {\cal H} \xi + ( {\cal H}' + 3 {\cal H}^2) C].
\label{ricciod}
\end{eqnarray}

\renewcommand{\theequation}{B.\arabic{equation}}
\setcounter{equation}{0}
\section{Decoupled equation for the Bardeen Potential}
Often in the course of this investigation the background equations have been 
used in order to reduce complicated expressions to simple ratios
of background functions. In order to give an example consider the equation
for the Bardeen potential obtained in Eq. (\ref{pseq}):
\begin{equation}
\Psi'' + [3 {\cal H} - 2 \frac{\varphi''}{\varphi'}] \Psi' + 
[ 4 {\cal H}' - 4 {\cal H} \frac{\varphi''}{\varphi'}] \Psi  - 
\partial_{\alpha}\partial^{\alpha} \Psi=0.
\label{psa1}
\end{equation}
To eliminate the first derivative term is trivial since it suffice to redefine 
$\Psi$ as 
\begin{equation}
\Phi = \frac{a^{3/2}}{\phi'} \Psi.
\end{equation}
Then, the equation for $\Phi$ will be 
\begin{equation}
\Phi'' - \partial_{\alpha}\partial^{\alpha} \Phi 
+ \Phi\biggl[ \frac{5}{2}{\cal H}'+ \frac{\varphi'''}{\varphi'}
 - \frac{9}{4} {\cal H}^2
-{\cal H} \frac{\varphi''}{\varphi'} 
- 2 \biggl( \frac{\varphi''}{\varphi'}\biggr)^2 \biggr]=0.
\label{inteq}
\end{equation}

It will now be shown that the quantity in squared bracket appearing in 
Eq. (\ref{inteq}) equals exactly $-z(z^{-1})''$ where $z= a^{3/2} 
\varphi'/{\cal H}$.
Notice, in fact, that 
\begin{equation}
-z(z^{-1})'' = - \frac{9}{4} {\cal H}^2 + \frac{9}{2} {\cal H}' 
 + \frac{\varphi'''}{\varphi'} 
- 2 \biggl(\frac{\varphi''}{\varphi'}\biggr)^2 - 
2 \frac{{\cal H}'}{\cal H} \frac{\varphi''}{\varphi'} 
- 3 {\cal H} \frac{\varphi''}{\varphi'}
-\frac{{\cal H}''}{\cal H}.
\label{inteq2}
\end{equation}
Recalling now, from Eqs. (\ref{c1})--(\ref{c2}),
 that ${\varphi'}^2 = 6 ({\cal H}^2 - {\cal H}')$
we can obtain (taking the derivative with respect to $w$):
\begin{equation}
\frac{{\cal H}''}{{\cal H}} = 2 {\cal H}' 
- \frac{\varphi'' \varphi'}{3 {\cal H}}.
\label{inteq3}
\end{equation}
Inserting Eq. (\ref{inteq3}) into Eq. (\ref{inteq2}) we get
\begin{equation}
-z(z^{-1})''= - \frac{9}{4} {\cal H}^2 + \frac{\varphi'''}{\varphi'}
+ \frac{5}{2} {\cal H}' 
 - 2 \biggl(\frac{\varphi''}{\varphi'}\biggr)^2 +
\frac{\varphi''}{\varphi'} \biggl( 2 \frac{{\cal H}'}{\cal H} - 3 {\cal H} 
+ \frac{{\varphi'}^2}{3 {\cal H}} \biggr)
\label{last}
\end{equation}
Using  ${\varphi'}^2 = 6 ({\cal H}^2 - {\cal H}')$ in the last bracket of 
Eq. (\ref{last}), we find 
that $-z(z^{-1})''$ is exactly what appears in the squared bracket of 
Eq. (\ref{inteq}). 
Similar algebraic manipulations are involved in the derivation 
of all the other decoupled equations discussed in the bulk of 
the paper.

\end{appendix}

\end{document}